\documentclass[12pt,twocolumn]{iopart}

\usepackage{graphicx}
\usepackage{dcolumn}
\usepackage{bm}
\usepackage{epsfig}

\begin{document}

\title{Protecting Superconducting Qubits with Universal Quantum Degeneracy Point}

\author{X.-H. Deng, Y. Hu, and Lin Tian}

\address{5200 North Lake Road, University of California, Merced, CA 95343, USA}

\ead{ltian@ucmerced.edu}

\begin{abstract}
Low-frequency noise can induce serious decoherence in superconducting qubits. Due to its diverse physical origin, such noise can couple with the qubits either as transverse or as longitudinal noise. Here, we present a universal quantum degeneracy point approach that can protect an encoded qubit from arbitrary low-frequency noise. We further show that universal quantum logic gates can be performed on the encoded qubits with high fidelity. The proposed scheme can be readily implemented with superconducting flux qubits or with a qubit coupling with a superconducting resonator. Meanwhile, the scheme is also robust against small parameter spreads due to fabrication errors in the superconducting systems.
\end{abstract}

\pacs{85.25.-j, 03.67.Lx, 03.67.Pp, 03.65.Yz}

\maketitle

\section{Introduction\label{Sec Introduction}}
Low-frequency noise is ubiquitous in solid-state systems \cite{DuttaHorn, Weissman}. In the past, extensive efforts have been devoted to study the microscopic origin of low-frequency noise in Josephson junction devices \cite{NECCharge1f, FluxQubitnoise, FluxQubitCCYu2012, Oliver2012, YYu2011, GunnarssonSST2013, FaoroPRL2008, Koch, FluxGeometry, PhaseQubit1fCC, PhaseQubit1fTemp, SQUIDDephasing, Zorin}. Recent theoretical and experimental efforts suggested that one source of low-frequency noise is the spurious two-level system fluctuators in the substrate or in the oxide layers of the Josephson junctions \cite{ShnirmanPRL2005, PhaseQubitTLS2, CCYUTLS, SYHan2012, Oliver2012dd}. In Ref. \cite{FluxQubitBylander2011}, the noise power spectral density of a superconducting flux qubit was thoroughly characterized by dynamical decoupling technique, which clearly demonstrates the low-frequency nature of the qubit noise. 

Decoherence due to low-frequency noise is one of the major hurdles in implementing fault-tolerant quantum computing in superconducting qubits \cite{SchonReview, YouReview, ClarkeReview}. To protect superconducting qubits from the decoherence caused by low-frequency noise, various approaches have been developed during the past few years, including dynamic control approach, quantum degeneracy point approach, using continuous measurement to calibrate the noise, and designing novel quantum circuits and materials \cite{NECEcho, Vion,IthierPRB, MakhlinOptimal, NECFluxQubitq1f, TianPRL2007, transmon1, fluxonium, material}. Among these approaches, the quantum degeneracy point approach effectively protects the qubit from low-frequency noise that couples to the qubit only through off-diagonal matrix elements, i.e., transverse noise \cite{Vion, IthierPRB, MakhlinOptimal, NECFluxQubitq1f}. Qubit decoherence time can be increased by orders of magnitude by operating the qubit at the quantum degeneracy point which is also called the optimal point or the ``sweet spot''. At the degeneracy point, the first-order derivative of the qubit energy splitting over noise is zero due to the low-frequency nature of the noise. 

Meanwhile, low-frequency noise in solid-state systems has diverse physical origins and can couple to the qubit either as transverse or as  longitudinal noise. In contrast to transverse noise, longitudinal noise couples to the qubit through diagonal matrix elements and generates a shift of the energy splitting. The traditional quantum degeneracy point approach \cite{Vion,IthierPRB, MakhlinOptimal, NECFluxQubitq1f} only protects the qubit from transverse low-frequency noise and cannot protect the qubit from longitudinal low-frequency noise. In this work, we propose a universal quantum degeneracy point (UQDP) approach where an encoded qubit can be protected from arbitrary low-frequency noise. The physical qubits are subject to transverse and/or longitudinal low-frequency noises. The encoded qubit is constructed in a subspace where the low-frequency noise only couples to off-diagonal matrix elements of the encoded qubit and can be treated as an effective transverse noise. We also show that universal quantum logic gates can be implemented on the encoded qubits in this architecture. The proposed scheme and the coupling circuits can be readily implemented with the superconducting flux qubits or with a qubit coupling with a superconducting resonator. To test our analytical results, we perform numerically simulation of the quantum gates and find that high gate fidelity can be achieved. We also show that the proposed approach is robust again small fabrication errors in the parameters of the Josephson junctions.  Furthermore, the UQDP approach is a general scheme that is not restricted to the superconducting qubits. It can be readily applied to a number of other quantum systems, such as the ion trap system, to suppress the decoherence of the low-frequency noise \cite{iontrap}. The paper is organized as follows. In Sec.\ref{Sec Universal point}, we present the UQDP approach with two coupled physical qubits. The decoherence of the encoded qubit under arbitrary low-frequency noise is calculated. In Sec.\ref{Sec Gates}, we study the realization of universal quantum logic gates on the encoded qubits. The implementation of this scheme with superconducting flux qubit is also presented. In Sec.\ref{Sec simulation}, we present numerical simulation of the gate operations to test our analytical results. In Sec.\ref{Sec qubit-resonator}, we further show that the encoded qubit in the UQDP approach can be implemented using a qubit coupling with a superconducting resonator. Discussions on the effect of parameter spreads and comparison with the decoherence free subspace (DFS) approach \cite{LidarDFS} are presented in Sec.\ref{Sec Discussions}. The conclusions are given in Sec.\ref{Sec Conclusions}.

\section{Universal Quantum Degeneracy Point with Coupled Qubits\label{Sec Universal point}}
\subsection{Quantum degeneracy point \label{SubSec QDP}}
The basic idea of the quantum degeneracy point approach is to use finite energy splitting of a qubit to protect the qubit from transverse low-frequency noise. Consider a qubit coupling to low-frequency noise as 
\begin{equation}
H_{\mathrm{s}}=E_{z}\sigma _{z}+\delta V_{x}(t)\sigma _{x},
\label{Equa Single Sample}
\end{equation}
where $2 E_{z}$ is the energy splitting between the qubit states $|\uparrow\rangle$ and $|\downarrow\rangle$ and $\delta V_{x}(t)$ is the noise amplitude with $|\delta V_{x}(t)| \ll E_{z}$. The noise couples to the qubit via the $\sigma_x$ operator where only the off-diagonal matrix elements are nonzero, and is hence a transverse noise. We treat the noise $\delta V_{x}(t)$ as a classical noise for the simplicity of discussion, but our results can be applied to quantum noises. The low-frequency nature of the noise determines that it cannot resonantly excite the qubit between its two states due to the large energy splitting of the qubit.  Hence, the noise can be treated as static fluctuations. With a second-order perturbation approach, the above Hamiltonian can be approximated as 
\begin{equation}
H_{\mathrm{s}}\approx (E_{z}+\delta V_{x}^{2}(t)/2E_{z})\sigma _{z},
\label{Eq:approxHs}
\end{equation}
where the qubit adiabatically follows the time dependence of the noise via a second-order term $\delta V_{x}^{2}(t)/2E_{z}$. Qubit dephasing is determined by this second-order term and is significantly suppressed by a factor of $\sim \left\vert \delta V_{x}(t)/2E_{z}\right\vert^2$  \cite{Vion,IthierPRB, MakhlinOptimal, NECFluxQubitq1f}.

However, as we mentioned in Sec.\ref{Sec Introduction}, the qubit-noise coupling in real systems can be more general than that in
Eq.(\ref{Equa Single Sample}). In the following, we consider a model of arbitrary low-frequency noise $\sum_\alpha \delta V_{\alpha}(t)\sigma _{\alpha }$ which includes coupling to all three Pauli matrices of the qubit. We will show that an encoded subspace can be constructed in which this noise can be converted to an effective transverse noise.

\subsection{Encoded qubit \label{SubSec Logic Qubit}}
Consider two qubits with the Hamiltonian
\begin{equation}
H_{\mathrm{0}}=E_{z}(\sigma _{z1}+\sigma _{z2})+\sum_{\alpha }E_{m\alpha}\sigma _{\alpha 1}\sigma _{\alpha 2},  \label{Eq Coupled Qubit Bare}
\end{equation}
where the $\sigma _{\alpha j}$'s are the Pauli matrices of the $j$-th qubit and $ \alpha =x,y,z$. An encoded qubit to implement the UQDP approach can be constructed from the above system when the condition $E_{mx}\ne 0$ and/or $E_{my}\ne 0$ is satisfied. As an example, we choose
\begin{equation}
E_{my}=E_{mz}=0,\quad E_{mx}=E_{m},\label{constant}
\end{equation}
where only the $x$-coupling is nonzero. Let the low-frequency noise have the general form
\begin{equation}
V_{\mathrm{n}}=\sum_{\alpha j}\delta V_{\alpha j}(t)\sigma _{\alpha j},
\label{Eq:Vn}
\end{equation}
where $\delta V_{\alpha j}(t)$ accounts for the noise component coupling with the $\sigma_{\alpha j}$ operator of the $j$-th qubit. The total Hamiltonian including the noise is then $H_{\mathrm{en}}=H_{\mathrm{0}}+V_{\mathrm{n}}$.

The eigenstates of $H_{\mathrm{0}}$ with the coupling given in Eq.(\ref{constant}) can be derived as
\begin{equation}
\begin{array}{l}
|1\rangle =-\sin (\theta/2) |\downarrow\downarrow\rangle +\cos (\theta/2)
|\uparrow\uparrow\rangle, \\ 
|2\rangle =\cos (\theta/2) |\downarrow\downarrow\rangle +\sin (\theta/2)
|\uparrow\uparrow\rangle, \\ 
|3\rangle =\left( -|\downarrow\uparrow\rangle +|\uparrow\downarrow\rangle
\right) /\sqrt{2}, \\ 
|4\rangle =\left( |\downarrow\uparrow\rangle +|\uparrow\downarrow\rangle
\right) /\sqrt{2},
\end{array}
\label{Eq Eigen}
\end{equation}
in terms of the $\vert \uparrow\rangle,\vert\downarrow\rangle$ states of the physical qubits with $\cos\theta=2E_z/\sqrt{4E_z^2+E_m^2}$. The corresponding eigenenergies are $E_1=-\sqrt{4E_z^2+E_m^2}$, $E_2=\sqrt{4E_z^2+E_m^2}$, $E_3=-E_m$, and $E_4=E_m$. The Pauli matrices of the physical qubits can be written in the eigenbasis. For example,
\begin{equation}
\sigma _{z1}=\left[ 
\begin{array}{cccc}
-\cos \theta & -\sin \theta & 0 & 0 \\ 
-\sin \theta & \cos \theta & 0 & 0 \\ 
0 & 0 & 0 & -1 \\ 
0 & 0 & -1 & 0
\end{array}
\right]  \label{Eq Z1}
\end{equation}
\begin{equation}
\sigma _{y2}=i\left[ 
\begin{array}{cccc}
0 & 0 & \cos \phi & \sin \phi \\ 
0 & 0 & \sin \phi & -\cos \phi \\ 
-\cos \phi & -\sin \phi & 0 & 0 \\ 
-\sin \phi & \cos \phi & 0 & 0%
\end{array}
\right]  \label{Eq Y2}
\end{equation}
with $\phi=\theta/2+\pi/4$. An interesting observation is that all the diagonal matrix elements of the Pauli matrices in the subspace spanned by $\left\{ |3\rangle ,|4\rangle \right\} $ are zero, i.e., $\left\langle 3\right\vert \sigma _{\alpha j}\left\vert 3\right\rangle =\left\langle 4\right\vert \sigma _{\alpha j}\left\vert 4\right\rangle =0$ for all $\alpha$ and $j$.  Furthermore, the only nonzero off-diagonal matrix elements in this subspace are $\langle 3|\sigma _{zj}|4\rangle $ and their conjugate elements; whereas 
\begin{equation}
\left\langle 3\right\vert \sigma _{xj}\left\vert 4\right\rangle
=\left\langle 4\right\vert \sigma _{xj}\left\vert 3\right\rangle
=\left\langle 3\right\vert \sigma _{yj}\left\vert 4\right\rangle
=\left\langle 4\right\vert \sigma _{xj}\left\vert 3\right\rangle =0.\label{Eq odme}
\end{equation}
Note that this result is valid for the general case of $E_{my},E_{mx}\neq 0$. Hence, $V_{n}$ only generates off-diagonal matrix elements in the subspace of $\left\{ |3\rangle ,|4\rangle \right\} $. These matrix elements, however, cannot induce strong transition between $|3\rangle $ and $|4\rangle $ due to the low-frequency nature of the noise and the finite energy splitting $2E_{m}$ between these two states. Similarly, the noise terms cannot induce strong transition between states in the subspace of $\left\{ |3\rangle ,|4\rangle \right\} $ and states in the subspace of $\left\{ |1\rangle ,|2\rangle\right\}$. Hence, we can select the subspace $\left\{ |3\rangle ,|4\rangle\right\} $ to form an encoded qubit that is protected from the low-frequency noise. We also define the parameter values where such encoded state exists as the universal quantum degeneracy point (UQDP). At a UQDP, the subspace $\left\{ |3\rangle,|4\rangle \right\} $ couples to the noise in a transverse way and suffers only quadratic dephasing from the low-frequency noise. In addition, the leakage from the encoded subspace to the redundant space of $\left\{ |1\rangle ,|2\rangle\right\} $ due to the low-frequency noise is also prohibited to the lowest order. The nonzero matrix elements of $\sigma _{xj}$ and $\sigma _{yj}$ (noise components $\delta V_{xj}(t)$ and $\delta V_{yj}(t)$) only induce virtual transitions between $\left\{ |3\rangle ,|4\rangle \right\} $ and $\left\{ |1\rangle ,|2\rangle \right\} $. As discussed above, UQDP exists under the condition $E_{mx}\ne 0$ and/or $E_{my}\ne 0$ which spans over a curve or a surface in the parameter space, not just a point. 

\subsection{Dephasing of encoded qubit \label{SubSec dephasing}}
The dephasing of the encoded qubit by arbitrary low-frequency noise can be calculated using a perturbation approach. Without loss of generality, we assume that the noise contains only a $x$-component $\delta V_{xj}(t)$ and a $z$-component $\delta V_{zj}(t)$ both of which are gaussian $1/f$ noises with the spectra 
\begin{equation}
S_{xj}(\omega )=\int_{-\infty }^{\infty }\left\langle \delta V_{xj}(t)\delta V_{xj}(0)\right\rangle \frac{e^{i\omega t}dt}{2\pi }=\frac{A^{2}\cos ^{2}\eta
}{\omega},  \label{Sx}
\end{equation}
\begin{equation}
S_{zj}(\omega )=\int_{-\infty }^{\infty }\left\langle \delta V_{zj}(t)\delta V_{zj}(0)\right\rangle \frac{e^{i\omega t}dt}{2\pi }=\frac{A^{2}\sin ^{2}\eta
}{\omega},  \label{Sz}
\end{equation}
where $\eta$ is the noise power angle that describes the distribution of the noise between the $x$- and $z$- components and $A^2$ is the total noise power. At $\eta=0$, the noise only contains a $x$-component and is a transverse noise for the physical qubits; at $\eta=\pi/2$, the noise contains a $z$-component and is a longitudinal noise for the physical qubits.

Define the Pauli operators for the encoded qubit as $X_{\sigma}=|3\rangle \left\langle 4\right\vert +|4\rangle \left\langle 3\right\vert $, $ Y_{\sigma}=-i|3\rangle \left\langle 4\right\vert +i|4\rangle \left\langle 3\right\vert $, and $Z_{\sigma}=|3\rangle \left\langle 3\right\vert -|4\rangle \left\langle 4\right\vert $. In the encoded subspace, the total Hamiltonian including the low-frequency noise can be approximated as
\begin{equation}
\displaystyle
H_{\mathrm{en}} = -E_{m} Z_{\sigma} +  \frac{E_{m}  (\delta V_{x1}^2(t) + \delta V_{x2}^{2}(t))}{2 E_{z}^{2}} Z_{\sigma} \label{eq:Hen}
- \frac{ (\delta V_{z1}-\delta V_{z2})^{2}(t)}{2E_{m}} Z_{\sigma} \label{eq:Hen}
\end{equation}
to the second order of the noise. The dephasing of the encoded qubit by arbitrary low-frequency noise only involves quadratic terms, similar to that in Eq.(\ref{Eq:approxHs}). 

To calculate the dephasing of the encoded qubit quantitively, we use the analytical results in Ref. \cite{MakhlinOptimal}. Assume that the noise components are not correlated with each other so that the cross terms involve the products of different $\delta V_{\alpha i}$ components can be dropped. We choose the following practical parameters for the superconducting systems: $ E_{z}/2\pi \hbar =5\,\mathrm{GHz}$, $A=2\times 10^{-4}E_{z}$, and an infrared cutoff for the $1/f$ noise $\omega _{ir}/2\pi =1\,\mathrm{Hz}$. For a superconducting flux qubit, given a flux noise $\delta \Phi (t)=\delta f (t) \Phi_{0}$ in the qubit loop, the noise amplitude can be written as $\delta V_{\alpha j}(t)=r_{1}E_{J} \delta f(t)$, where $E_{J}/2\pi\hbar=200\,\mathrm{GHz}$ is the Josephson energy of the loop junctions and $r_{1}\sim 5$ with typical parameters \cite{OrlandoPRB1999}. Using this relation, the power spectral density of the flux noise under our noise power is $(10^{-6}\Phi_{0})^{2}/\omega$, comparable the experimental results in Ref. \cite{FluxQubitBylander2011}.

\begin{figure}
\begin{center}
\includegraphics[width=7.5cm,clip]{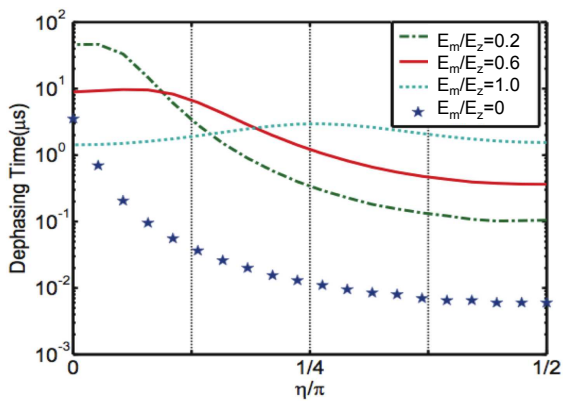}
\caption{Dephasing time of the bare qubit (star) and the encoded qubit versus the noise power angle $\eta$. The values of $E_{m}/E_{z}$ for corresponding curves are given in the inset. The spectra of the low-frequency noise are given in Eq.(\ref{Sx},\ref{Sz}).}
\label{Fig FIDLaw}
\end{center}
\end{figure}

The dephasing time for an encoded qubit is plotted in Fig.\ref{Fig FIDLaw} at selected ratios of $E_{m}/E_{z}$. For zero or small ratio, e.g., $E_m/E_z=0.2$, the dephasing time decreases rapidly by a few orders of magnitude as the noise changes from transverse to longitudinal noise (as $\eta$ increases). For a sizable ratio, e.g. $E_{m}/E_{z}=0.6$, the decoherence time varies slowly over the whole range of the distribution angle $\eta $.  We also find that given a fixed noise power $A^2$, the dephasing time reaches a maximum at $E_m/E_z\approx \sin\eta/\cos\eta$. For example, the maximal dephasing time at $E_{m}/E_{z}=1$ is reached at $\eta=\pi/4$. This can be explained by the noise terms in Eq.(\ref{eq:Hen}) ---  the maximum in the dephasing time occurs when the second and the third term in Eq.(\ref{eq:Hen}) make equal contribution to dephasing. 
\small\normalsize

Note that the superconducting qubits are also subject to decoherence by high-frequency noise, e.g., Ohmic noise in the electrical circuits, which induces relaxation of the qubits. Assume that the relaxation rate of the physical qubits is $\gamma_{h}$. It can be readily shown that the relaxation rate of the encode qubit due to the high-frequency noise is of the same order of magnitude as $\gamma_{h}$. 

\section{Universal Quantum Logic Gates \label{Sec Gates}}
In Sec.\ref{Sec Universal point}, we showed that the dephasing of the encoded qubit is dominated by quadratic terms of the low-frequency noise. The encoded qubit can thus be a robust quantum memory to store quantum information. Here, we will further show that universal quantum logic gates can be implemented on the encoded qubits with high fidelity. 

\subsection{Single-qubit gates \label{SubSec single-qubit gates}}
Single-qubit gates on an encoded qubit can be performed by manipulating either the energy of individual physical qubit or the coupling between the qubits. Projected to the encoded subspace, the Hamiltonian $H_{\mathrm{0}}$ becomes $P_{e}H_{\mathrm{0}}P_{e}=-E_m Z_{\sigma}$ and the operator $\sigma _{z1}$ becomes $P_{e}\sigma _{z1}P_{e}=-X_{\sigma}$, where $P_{e}$ is the projection operator. Here $\sigma _{z1}$ generates no coupling between the encoded subspace and subspace spanned by $\left\{ |1\rangle, |2\rangle \right\}$, as is shown in Eq. (\ref{Eq Z1}). Hence, driving the first physical qubit with a pulse $H_{X}=2\lambda \cos(2E_{m} t/\hbar)\sigma _{z1}$, we derive an effective Hamiltonian $H_{X}^{rot}=-\lambda X_{\sigma}$ in the rotating frame. After a duration of $\theta \hbar/2\lambda $, the $X$-rotation $U_{X}(\theta)=\exp (i\theta X_{\sigma}/2)$ can be implemented on the encoded qubit. Similarly, when projected to the encoded subspace, the operator product $\sigma _{x1}\sigma _{x2}$ becomes $P_{e}\sigma _{x1}\sigma _{x2}P_{e}=- Z_{\sigma}$, which generates a $Z$-rotation $U_{Z}$ on the encoded qubit. These two operations form a complete $SU(2)$ generator set that can implement arbitrary single-qubit gate on the encoded qubit.

\subsection{Two-qubit gates \label{SubSec two-qubit gates}}
\begin{figure}
\begin{center}
\includegraphics[width=10cm,clip]{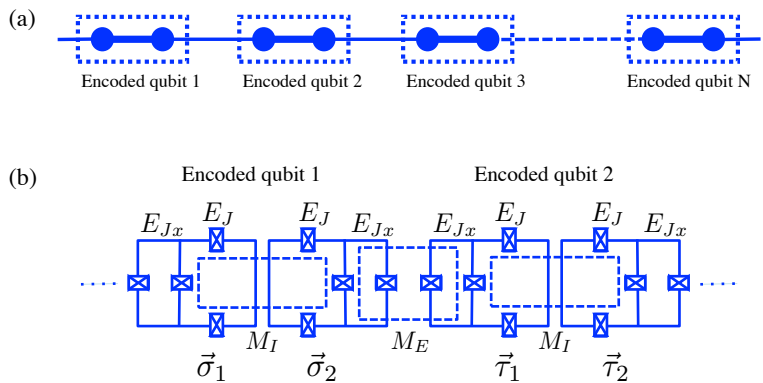}
\caption{(a) Illustration of the encoded qubits. The physical qubits are represented by solid circles and the horizontal bars indicate their coupling. Each encoded qubit is enclosed in a dashed box. The thin bars indicate the coupling between physical qubits in neighboring encoded qubits. (b) Implementation of the encoded qubits with superconducting flux qubits.}
\label{Fig Coupled Logic Qubit}
\end{center}
\end{figure}

Controlled quantum logic gates on the encoded qubits can be implemented by coupling the physical qubits in two encoded qubits, as is illustrated in Fig.\ref{Fig Coupled Logic Qubit} (a). We use $\sigma_{\alpha j}$ ($\tau_{\alpha j}$) as the Pauli matrices for the physical qubits in the first (second) encoded qubit. Assume  that the physical qubit $\vec{\sigma}_{2}$ in the first encoded qubit is coupled with the physical qubit $\vec{\tau}_{1}$ in the second encoded qubit with $H_{C}= \lambda _{c} \sigma_{z2}\tau _{z1}$. Given that $P_{e}\sigma _{z2}P_{e}=X_{\sigma}$ and $P_{e}\tau _{z1}P_{e}=-X_{\tau}$, this coupling can be projected to the encoded subspace as $H_{C}= -\lambda _{c} X_{\sigma}X_{\tau}$. We can use this coupling to perform two-qubit operations. By adjusting the energy splittings of the encoded qubits to have $E_{m\sigma}=E_{m\tau}$, where $E_{m\sigma}$ and $E_{m\tau}$ are energy splittings of the encoded qubits, we have in the interaction picture
\begin{equation}
H_{C}^{\mathrm{(I)}}=-\frac{\lambda _{c}}{2}(X_{\sigma}X_{\tau}+Y_{\sigma}Y_{\tau}),\label{cgate1}
\end{equation}
after neglecting the fast oscillating terms. This coupling can be used to perform swap gate or $\sqrt{\mathrm{swap}}$ gate between the encoded qubits. 

We can also design a tunable coupling between two qubits with \cite{YXLiuPRL2006QubitCoupling, Sembat1, BertetPRB2006, NiskanenPRB2006, NiskanenScience2007, tunableCCLtian, tunableCC}
\begin{equation}
H_{C}=2\lambda _{c}\cos (2(E_{m\sigma}-E_{m\tau})t/\hbar )\sigma_{z2}\tau _{z1},  \label{Eq:cgate2}
\end{equation}
which again gives us the effective coupling in Eq.(\ref{cgate1}) in the interaction picture and does not require $E_{m\sigma}=E_{m\tau}$. 

The single-qubit and two-qubit gates form a universal set of quantum logic operations for the encoded qubits. We can also design various architectures to connect the encoded qubits. For example, in Fig.\ref{Fig Coupled Logic Qubit} (a), the second (right hand side) physical qubit of the $n$th encoded qubit is coupled to the first (left hand side) physical qubit of the $(n+1)$th encoded qubit. The system can be viewed as a one-dimensional chain of qubits with nearest neighbor coupling. Universal quantum computation can be implemented in this architecture. 
\small\normalsize

\subsection{Implementation with superconducting flux qubit \label{Sec implementation}}
The UQDP approach can be implemented in various systems given the diversity of the superconducting qubits. Here, we present an implementation with the superconducting flux qubits. As illustrated in Fig.\ref{Fig Coupled Logic Qubit} (b), the encoded qubit is composed of two flux qubits labeled as $\vec{\sigma}_{1,2}$ (or $\vec{\tau}_{1,2}$) coupling via mutual inductance $M_{I}$ between the main qubit loops. The flux qubit is made of four Josephson junctions: two loop junctions with Josephson energy $E_{J}$ and two SQUID junctions with Josephson energy $E_{Jx}$. By varying the magnetic flux $\Phi_{loop}$ inside the main loop, the operating point can be adjusted; and by varying the magnetic flux $\Phi_{sq}$ inside the SQUID loop between the two $E_{Jx}$-junctions, the quantum tunneling between the persistent-current states can be adjusted \cite{MooijScience1999, OrlandoPRB1999}. We choose the operating point to be $(\Phi_{loop}+\Phi_{sq}/2)/\Phi_{0}=1/2$ with, e.g., $\Phi_{loop}/\Phi_{0}=\Phi_{sq}/\Phi_{0}=1/3$, which is the degeneracy point for the flux qubit. The qubit Hamiltonian is then $H_{\sigma i}=E_{z}\sigma_{zi}$, where $E_{z}$ is the quantum tunneling between the two persistent-current states and the eigenstates are equal superpositions of the persistent-current states. Under the eigenbasis, the circulating current operator of the qubit can be written as $I_{cir}\sigma_{xi}$. The mutual inductance coupling between the main loops induces a shift of the magnetic flux $\Phi_{loop}$ in the main loops, which shifts the qubit energy \cite{MajerPRL2005}. The interaction energy has the form of $H_{MI}=E_{m}\sigma_{x1}\sigma_{x2}$, where the coupling $E_{m}$ is determined by the mutual inductance and the loop currents as was discussed in detail in Ref. \cite{OrlandoPRB1999}. The total Hamiltonian hence has the form of Eq.(\ref{Eq Coupled Qubit Bare}) with $E_{mx}\ne 0$ and $E_{my}=E_{mz}=0$. With practical circuit parameters, the quantum tunneling can reach a few gigahertz \cite{FluxQubitBylander2011} and the mutual inductance coupling can reach gigahertz \cite{NiskanenScience2007}, which satisfy the operating conditions of the encoded qubits.

The X-rotation $U_{X}(\theta)$ on single qubit can be realized by applying a magnetic flux pulse at the microwave frequency $2E_{m}$ to the SQUID loop of qubit $\vec{\sigma}_{1}$. This pulse generates a $\sigma_{z1}$ operation which can be projected to the encoded space to generate the $U_{X}$ operation. The Z-rotation $U_{Z}(\theta)$ can be realized by using a tunable coupling between the physical qubits to generate the term $ H_{Z}=\delta E_{m} \sigma_{x1} \sigma_{x2}$ which can be projected to the encoded subspace as $H_{Z}=- \delta E_{m} Z$. 

Neighboring encoded qubits couple via the mutual inductance between the SQUID loops of the physical qubits. In Fig.\ref{Fig Coupled Logic Qubit} (b), the SQUID loop of the physical qubit $\vec{\sigma}_{2}$ inductively couples with the SQUID loop of the physical qubit $\vec{\tau}_{1}$ via mutual inductance $M_{E}$. The circulating current in the SQUID loops generates two effects on the other qubit. First, it shifts the magnetic flux $\Phi_{sq}$ in the SQUID loop of the other qubit and hence modifies the quantum tunneling. This gives us a coupling of the form $\lambda_{c}\sigma_{z2}\tau_{z1}$ which can be used to implement controlled quantum logic gates. Second, the circulating current in the SQUID loop shifts the total magnetic flux $\Phi_{loop}+\Phi_{sq}/2$ of the qubit, which modifies the flux bias of the persistent-current states and gives us a coupling of the form $\lambda_{c2}\sigma_{z2}\tau_{x1}$. This coupling has no effect on the encoded subspace and can be neglected in our discussion. Combining the two-qubit coupling and single qubit operations, universal gate operations on the encoded qubits can be realized with superconducting flux qubits. 
 
\subsection{State preparation and detection\label{SubSec prep detect}}
To perform the gate operations, the system needs to be prepared in a proper initial state in the encoded subspace. First, the coupled-qubit  system can be relaxed to its ground state, i.e., state $|1\rangle$ in Eq.(\ref{Eq Eigen}), by thermalization. Then, an ac pulse is applied with
\begin{equation}
H_{\mathrm{prep}}=2\lambda _{p}\cos \left[ (E_{3}-E_{1})t/\hbar\right] \sigma _{x1}
\end{equation}
to the qubit $\vec{\sigma}_{1}$. Because $\sigma _{x1}$ has the nonzero matrix element $\langle 3| \sigma _{x1}|1\rangle=-\sin (\theta/2+\pi/4)$, this pulse generates a Rabi oscillation between states $|1\rangle$ and $|3\rangle$ when neglecting fast oscillating terms. After a duration of $\pi /2\lambda_{p}\sin(\theta/2+\pi/4)$, the state becomes $|3\rangle$ which is the lower eigenstate in the encoded subspace.

State detection can be implemented by applying single-qubit rotation as well. To measure the probability of an encoded qubit in the state $|3\rangle$ or $|4\rangle$, apply the single-qubit gate $\exp(-i\pi X/4)$. This operation converts state $|3\rangle$ to the state $|\uparrow\downarrow\rangle$ and converts state $|4\rangle $ to the state $|\downarrow\uparrow\rangle$. After this operation, we can measure the physical qubits to extract information of the encoded states. 

The above state preparation and detection schemes can be readily realized with superconducting flux qubit setup presented in Sec.\ref{Sec implementation}.

\section{Numerical Simulation \label{Sec simulation}}
To test our analytical results, we conduct numerical simulation of the quantum logic gates and derive the gate fidelity. In the simulation, the low-frequency noise is generated using a stochastic sequence given below in Eq.~(\ref{Vsimu}). The time evolution under the total Hamiltonian for the physical qubits during the gate operation is simulated with a fourth-order Runge-Kutta method. The simulation will be repeated several hundreds of times with a new sequence of randomly-generated low-frequency noise $V_n(t)$ each time. The gate fidelity is obtained by averaging over all the simulation runs.

We make the following assumptions about the low-frequency noise. Firstly, the noise couples with the physical qubits in the form of $\delta V_{xj}(t)\sigma _{xj}+\delta V_{zj}(t)\sigma _{zj}$ which contains both longitudinal and transverse noise and can be used to demonstrate the decoherence property of the encoded qubit.  Secondly, the noise has a $1/f$ spectrum with a noise power $A^{2}$. Thirdly, the distribution between the longitudinal and the transverse component of the noise is characterized by the noise power angle $\eta$ which is  defined below.  The $1/f$-noise can be generated using a stochastic sequence:
\begin{equation}
V_{\alpha j}(t)=\sum_{\omega =\omega _{ir}}^{\omega _{uv}}a_{\alpha j}(\omega )\cos (\omega t+\phi )\Delta \omega   \label{Vsimu}
\end{equation}
where $a_{\alpha j}$ obeys gaussian distribution with zero average. For transverse noise,
\begin{equation}
\langle a_{\alpha x}(\omega _{1})a_{\alpha x}(\omega _{2})\rangle =A^{2}\cos^{2}(\eta )\delta (\omega _{1}+\omega _{2})/\omega _{1},\label{aax}
\end{equation}
and for longitudinal noise
\begin{equation}
\langle a_{\alpha z}(\omega _{1})a_{\alpha z}(\omega _{2})\rangle =A^{2}\sin^{2}(\eta )\delta (\omega _{1}+\omega _{2})/\omega _{1}.  \label{aaz}
\end{equation}
Here discrete noise components with a separation of $\Delta \omega /2\pi =10^{-4}\,\textrm{MHz}$ are used to replace the continuous integral of the spectral density. The phase $\phi $ is a random number with a uniform distribution between $0$ and $2\pi $. Our parameters are $E_{z}/2\pi \hbar =5\,\mathrm{GHz}$, the infrared frequency limit $\omega _{ir}/2\pi =1\,\mathrm{Hz}$, the upper bound of the noise $\omega _{uv}/2\pi =0.1\,\mathrm{MHz}$, and the noise power $A/E_{z}=2\times 10^{-4}\,\mathrm{s/rad}$ \cite{FluxQubitBylander2011}.

\small\begin{figure}
\begin{center}
\includegraphics[width=10cm,clip]{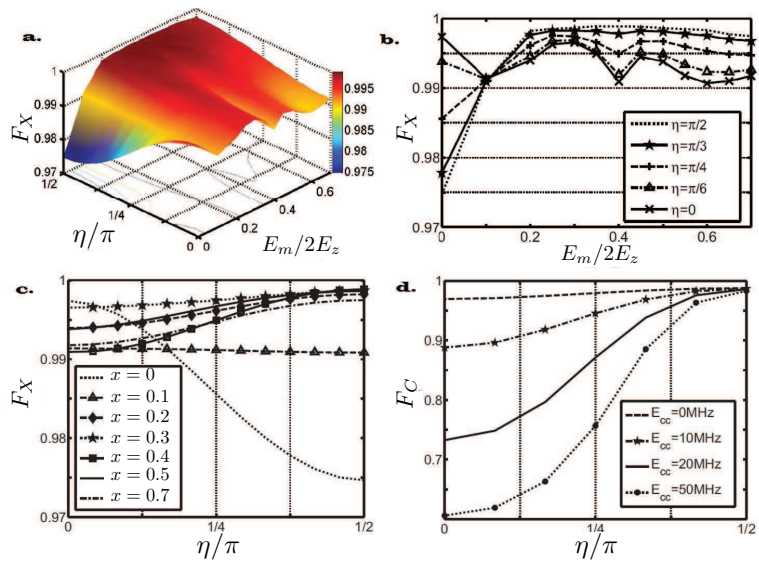}
\caption{(a) Gate fidelity $F_{X}$ for single-qubit gate $U_{X}(\protect\pi )$ versus power distribution angle $\eta$ and coupling $E_{m}/2E_{z}$. (b) $F_{X}$ versus $E_{m}/2E_{z}$ at selected $\eta$ values. (c) $F_{X}$ versus $\eta$ at selected $x=E_{m}/2E_{z}$ values. (d) Gate fidelity $F_{C}$ of two-qubit gate $U_{C}$ versus $\eta$ at selected $E_{cc}$ (in unit of $2\pi\hbar$) values.}
\label{FX3D}
\end{center}
\end{figure}\normalsize
For the single-qubit gate $U_{X}(\pi)$, the gate fidelity can be defined as \cite{NielsonFidelity} 
\begin{equation}
F_{X}=\frac{1}{2}+\frac{1}{12}\sum_{i=1,2,3}\mathrm{Tr}\left(U_{X}(\pi )\Sigma_{i}U_{X}^{\dag }(\pi )\varepsilon (\Sigma _{i})\right),\label{FX_def}
\end{equation}
where $\Sigma _{i}$ are the Pauli matrices of the encoded qubit with $\Sigma _{i}=X,\,Y,\,Z$ for $i=1,2,3$, respectively. The super-operator $\varepsilon (\Sigma _{i})=P_{e}\mathcal{L}(\Sigma_{i})P_{e}^{\dag }$ is the projection of the final state to the encoded subspace after applying the gate operation to the initial density matrix $\Sigma _{i}$. In the simulation, the constant $\lambda$ in $H_{X}$ is chosen as $\lambda /2\pi \hbar=300\,\mathrm{MHz}$. The gate fidelity $F_{X}$ of the operation $U_{X}(\pi )$ is plotted in Fig.\ref{FX3D} (a) - (c). For a sizable ratio of $E_{m}/2E_{z}$, where the encoded subspace is well protected from the low-frequency noise, the fidelity varies smoothly within a narrow regime as $\eta$ increases from $0$ (transverse noise) to $\pi/2$ (longitudinal noise). While for $E_{m}=0$, i.e., uncoupled qubits, the fidelity decreases quickly as $\eta$ increases. Meanwhile, for finite $\eta$ (sizable longitudinal component), the fidelity first increases quickly with $E_{m}$, but then becomes saturated and even shows oscillatory behavior as $E_{m}$ further increases. With $E_{m}\sim 0.4 E_{z}$, $F_{X}$ exceeds $0.997$ for $\eta \in [0,\pi /2]$. According to our numerical simulation, an optimal value of $E_{m}$ for a particular $\eta $ usually can be found within the range of $(0.4E_{z},\,E_{z})$.

For the two-qubit gate $U_{C}=\exp [i\pi (X_{1}X_{2}+Y_{1}Y_{2})/4]$, the gate fidelity can be defined as \cite{NielsonFidelity} 
\begin{equation}
F_{C}=\frac{1}{5}+\frac{1}{80}\sum_{i,j}\mathrm{Tr}(U_{C}(\Sigma _{i}\otimes \Omega _{j})U_{C}^{\dag }\varepsilon (\Sigma _{i}\otimes \Omega _{j})), \label{FC_def}
\end{equation}
where $\Omega _{j}$ are the Pauli matrices of the second encoded qubit with $\Omega_{j}=X,\,Y,\,Z$ for $j=1,2,3$ respectively. The super-operator $\varepsilon (\Sigma_{i}\otimes \Omega _{j})=P_{e}\mathcal{L}(\Sigma _{i}\otimes \Omega_{j})P_{e}^{\dag }$ is the projection of the final state to the encoded subspace after applying the quantum operation to the initial state $\Sigma _{i}\otimes \Omega_{j}$. To implement this gate,  we use the tunable coupling given in Eq.(\ref{Eq:cgate2}) plus an extra small term $E_{cc}\sigma_{x2}\tau_{x1}$ which describes the inductive coupling between the loop currents of $\vec{\sigma}_{2}$ and $\vec{\tau}_{1}$. By carefully designing the circuit, this term can be made much weaker than the coupling in Eq.(\ref{Eq:cgate2}) with $E_{cc}\ll\lambda_{c}$. We choose $E_{m\sigma}/2\pi \hbar =5\,\mathrm{GHz}$, $E_{m\tau}/2\pi \hbar =2\,\mathrm{GHz}$, and $\lambda _{c}/2\pi \hbar =300\,\mathrm{MHz}$. The tunable coupling is applied for a duration of $\pi \hbar /4\lambda _{c}$. The gate fidelity $F_{C}$ for this operation is plotted in Fig.\ref{FX3D} (d) at selected values of $E_{cc}$. It can be seen that for $E_{cc}=0$, $U_{C}$ can be accomplished with high fidelity and shows weak dependence on the noise power angle $\eta$. However, for larger $E_{cc}$, e.g., $E_{cc}/2\pi\hbar=50\,\mathrm{MHz}$, the gate fidelity decreases quickly with $\eta$. This shows the necessity to make this term small in the circuit design. 

\section{Universal Quantum Degeneracy Point with Qubit-Resonator System\label{Sec qubit-resonator}}
In the UQDP approach, the encoded subspace is designed to couple with low-frequency noise only through transverse couplings and the finite energy splitting between the eigenstates protects the encoded qubit from low-frequency noise. Here we show that UQDP can also be realized with a qubit coupling with a superconducting resonator. 

Consider a qubit coupling with a resonator mode in the Jaynes-Cummings model
\begin{equation}
H_{qr}=\hbar\omega_{0}a^{\dag}a + E_{z}\sigma_{z} + J \left(a^{\dag}\sigma_{-}+\sigma_{+}a\right),\label{JC model}
\end{equation}
where $a$ ($a^{\dag}$) is the annihilation (creation) operator of the resonator, $\omega_{0}$ is the resonator frequency, and $J$ is the magnitude of the qubit-resonator coupling. Assume $\hbar\omega_{0}=2E_{z}$. The eigenstates of this system include the ground state $|0\downarrow\rangle$ and the polariton doublets
\begin{equation}
|n\alpha\rangle=\left(|n\downarrow\rangle+(-1)^{\alpha}|(n-1)\uparrow\rangle\right)/\sqrt{2}\label{doublet}
\end{equation}
with integer $n$ ($n\ge1$) and $\alpha=0,1$. The eigenenergies of the polariton doubles are $E_{n\alpha}=\hbar\omega_{0}/2 + (-1)^{\alpha}J$ with an energy splitting $2J$ between the doublet states. The subspace of a doublet can be chosen as an encoded qubit.

Assume the low-frequency noise couples with the qubit as $V_{nq}=\sum_{j}\delta V_{qj}(t)\sigma_{j}$ and couples with the resonator as $V_{nr}=\delta V_{r}(t)a^{\dag} + h.c.$.  The effect of the low-frequency noise on the encoded qubit can be derived from the noise coupling operators. Using Eq.(\ref{doublet}), we have
\begin{equation}
\langle m\alpha |\sigma_{z}| n\beta \rangle=\delta_{m,n}(-1+(-1)^{\alpha+\beta})/2,
\end{equation}
where the only nonzero matrix elements are the off-diagonal elements at $m=n$ and $\alpha\neq\beta$. We also have 
\begin{equation}
\langle m\alpha |\sigma_{+}| n\beta \rangle=\delta_{m,n+1}(-1)^{\alpha}/2,
\end{equation}
\begin{equation}
\langle m\alpha |a^{\dag}| n\beta \rangle=\delta_{m,n+1}(\sqrt{n+1}+(-1)^{\alpha+\beta}\sqrt{n})/2,
\end{equation}
and similar results for their conjugate operators, where the only nonzero matrix elements are between states in adjacent doublets. Given these results, the low-frequency noise only induces off-diagonal matrix element in the encoded subspace of the polariton doublet, similar to that in Sec.\ref{Sec Universal point}. The encoded qubit in the qubit-resonator system is hence protected from the low-frequency noise. 

\section{Discussions \label{Sec Discussions}}
\subsection{Robustness under parameter spreads\label{Sec circuit error}}
In superconducting devices, the circuit parameters can have a fabrication error of the order of $5\%$ which is limited by the micro-fabrication technology. The energy splittings of the physical qubits in an encoded qubit thus cannot be exactly equal to each other. We introduce $a_{0}$ as the ratio between the energy splittings of the qubit $\vec{\sigma}_{2}$ and the qubit $\vec{\sigma}_{1}$ with $|a_0-1|\ll 1$. The Hamiltonian in Eq.(\ref{Eq Coupled Qubit Bare}) becomes
\begin{equation}
H_{\mathrm{e}}=E_{z}(\sigma _{z1}+a_{0}\sigma _{z2})+E_{m}
\sigma_{z1}\sigma _{z2}.  \label{Eq NonUniformHami}
\end{equation}
The states $\left\{ |3\rangle ,|4\rangle \right\}$ in Eq.~(\ref{Eq Eigen}) still span the encoded subspace, but the eigenstates are now rotated with
\begin{equation}
\begin{array}{lcl}
|3\rangle _{\mathrm{e}}=\cos \varphi |3\rangle +\sin \varphi |4\rangle , & 
\quad & E_{\mathrm{3e}}=-E_{\mathrm{e}}; \\ 
|4\rangle _{\mathrm{e}}=-\sin \varphi |3\rangle +\cos \varphi |4\rangle , & 
\quad & E_{\mathrm{4e}}=E_{\mathrm{e}};
\end{array}
\end{equation}
and the eigenenergies are shifted with $E_{\mathrm{e}}=\sqrt{(1-a_{0})^{2}E_{\mathrm{J}}^{2}+E_{\mathrm{m}}^{2}}$ and $\varphi =\sin ^{-1}[(a_{0}-1)E_{\mathrm{z}}/E_{\mathrm{e}}]/2$. It can be shown that the qubit operators $\sigma _{x1}$, $\sigma _{x2}$, $\sigma_{y1}$, and $\sigma _{y2}$ contain only nonzero matrix elements that connect $\left\{ |1\rangle _{\mathrm{e}} ,|2\rangle _{\mathrm{e}} \right\}$ and $\left\{ |3\rangle _{\mathrm{e}} ,|4\rangle _{\mathrm{e}} \right\}$. However, the operators $\sigma _{z1}$ and $\sigma _{z2}$ contain diagonal terms 
\begin{equation}
\begin{array}{l}
_{\mathrm{e}}\langle 3|\sigma _{z1}|3\rangle _{\mathrm{e}} = -_{\mathrm{e}}\langle 4|\sigma _{z1}|4\rangle _{\mathrm{e}}=\frac{-(1-a_{0})E_{\mathrm{z}} }{E_{\mathrm{e}}}, \\ 
_{\mathrm{e}}\langle 3|\sigma _{z2}|3\rangle _{\mathrm{e}} = -_{\mathrm{e}}\langle 4|\sigma _{z2}|4\rangle _{\mathrm{e}}=\frac{(1-a_{0})E_{\mathrm{z}}}{E_{\mathrm{e}}},%
\end{array}
\label{extra34}
\end{equation}
which generate residual longitudinal noise on the encoded qubit. This residual noise is
\begin{equation}
V_{\mathrm{res}}= (a_{0}-1)(E_{\mathrm{z}}/E_{\mathrm{e}})
(\delta V_{z1}-\delta V_{z2})Z,  \label{Eq ResidualCoupling}
\end{equation}
which induces dephasing on the encoded qubit with first-order noise terms. However, the residual noise contains a small ratio $|a_0-1|$ which is often less than $5\%$. Dephasing due to this residue noise is then reduced by a factor $|a_0-1|^2< 2\times10^{-3}$, which reduces the dephasing rate to be comparable to or even lower than the quadratic dephasing rate due to the transverse noise in Eq.~(\ref{eq:Hen}).

In previous sections, UQDP is studied under the coupling $E_{mx}=E_{m},\,E_{my}=E_{mz}=0$. Note that UQDP works for a broad range of couplings given in Eq.(\ref{Eq Coupled Qubit Bare}) as far as the condition $E_{mx}\ne 0$ and/or $E_{my}\ne 0$ is satisfied. For example, we can choose $E_{mx}=E_{m},\, E_{my}=E_{mz}=b_0 E_m$ with finite $b_0$. Under this coupling, the subspace $\{|3,4\rangle\}$ still forms an encoded qubit. The eigenenergies are $E_{3} =-E_{m}-2b_{0}E_{m}$ and $E_{4} = E_{m}$, including an energy shift of $2b_0$. This encoded qubit is protected against first-order dephasing due to arbitrary low-frequency noise. This observation shows that the UQDP approach can be applied to a variety of superconducting qubits as well as other quantum systems such as the ion trap system \cite{iontrap}.

\subsection{Comparison with decoherence free subspace approach\label{Sec DFS}}
The UQDP approach is different from the decoherence free subspace (DFS) approach that has been widely studied \cite{LidarDFS}. The DFS approach protects the qubits against spatially correlated noises by choosing a subspace that is immune to such noise, i.e., the decoherence is suppressed by the symmetry of the noise. While in our scheme, we exploit the energy splitting of the encoded qubit and the low-frequency nature of the noise (the noise cannot resonantly excite transitions between states with large energy splitting) to protect the quantum state. 

\section{Conclusions \label{Sec Conclusions}}
In conclusion, we propose a UQDP approach to protect the superconducting qubits from arbitrary low-frequency noise. Using two coupled qubits to form an encoded qubit, we find a subspace where the low-frequency noise only causes dephasing in quadratic order. We show that universal quantum logic gates can be performed on the encoded qubits with high fidelity. The approach is robust against parameter spreads due to fabrication errors. We also show that this scheme can be realized with a qubit coupling with a superconducting resonator. Furthermore, this approach can be applied to a broad range of physical systems to reduce the effect of low-frequency noise.

\section{Acknowledgments}
This work is supported by the National Science Foundation under Grant No. NSF-CCF-0916303 and NSF-DMR-0956064. XHD is partially supported by a scholarship from China.

\end{document}